\documentclass[11pt]{article}
\usepackage{jcappub}

\usepackage{amsmath}
\usepackage{amssymb}
\usepackage{graphicx}
\usepackage{epsfig}
\usepackage{caption}
\usepackage{subcaption}

\begin{document}

\title{Probing the Scale of ALP Interactions with Fermi Blazars}

\author{Rebecca Reesman$^{1,2}$}
\emailAdd{rreesman@physics.osu.edu}
\author{T.P. Walker$^{1,2,3}$}
\affiliation{$^{1}$Department of Physics, The Ohio State University, Columbus, OH 43210}
\affiliation{$^{2}$Center for Cosmology and Astro-Particle Physics, The Ohio State University, Columbus, OH 43210} 
\affiliation{$^{3}$Department of Astronomy, The Ohio State University, Columbus, OH 43210}

\abstract{
Gamma-rays from cosmological sources contain information about gamma-ray interactions. Standard model and non-standard model photon interactions along the path between the source and the observer can lead to changes in the energy or state of the photons, which in turn alters the observed energy spectrum of the source. In general, these interactions are a function of photon energy as well as source distance.  Here we show how existing high energy gamma-ray observations of blazars can be used to constrain the coupling of axion-like-particles (ALPs) to the photon.  The same ALP-photon coupling that has been invoked to explain the observations of TeV blazars beyond their pair production horizon is shown to have an effect of the data set of  \textit{Fermi} blazars.}

\keywords{axions, gamma-ray theory }

\maketitle

\section{Introduction} 

Cosmological gamma-rays are an ideal probe of relatively rare photon interactions. In addition to expected standard model interactions, like those with the extragalactic background light (EBL) through pair production, any additional interaction which alters the energy or state of the photon can in principle be probed.  The process of constraining gamma-ray interactions beyond the standard model depends on knowledge of the source gamma-ray spectra, the quality of rapidly growing data sets of high energy gamma-ray observations, and a sufficiently well understood modeling of the EBL. As a test case, we consider a group of gamma-ray sources at approximately the same distance, using the  high redhift ($z \sim 1$) portion of the {\textit Fermi} blazar catalogue \cite{oai:arXiv.org:1108.1435} as an example. In addition, we discuss a similar set of TeV blazars clustered at $z\sim 0.1$ (see references in \cite{Reesman:2013jra}) . We expect blazars at the same distance to have their spectra be attenuated in a similar way by the EBL (for a recent review see \cite{Costamante:2013sva}).  If we assume a well-determined model for the EBL then, in principle,  a comparison of the observed spectra to the spectra at the source can probe new photon couplings. This requires some assumptions about the intrinsic spectra, namely that they are a power  law. In practice, we then use the residuals between the observed (as measured by the {\it Fermi} blazar data set) and expected EBL attenuation as an example of the range of constraints that can be placed on rare photon interactions. There have been observations of distant gamma-ray sources which suggest that the Universe is less opaque to gamma-rays than expected \cite{Aharonian:2005gh, Aleksic:2011hr, Acciari:2009zy, Aliu:2008rg}. Rare photon couplings have recently been proposed to explain blazar spectra \cite{SanchezConde:2009wu, Wouters:2013iya, Tavecchio:2012um, Wouters:2012qd}, leading  to general questions about novel couplings of cosmological photons, which we show could be probed by the \textit{Fermi} blazars.

Specifically, we examine  the mixing of the photon with a pseudoscalar axion-like particle (ALP).  ALPs couple to photons via a two-photon vertex, just as the axion\cite{PhysRevLett.38.1440}  does.  Unlike the axion, which has a definite relationship between this coupling and its mass, the ALP's mass is independent of coupling \cite{Kim:1998va, Mirizzi:2006zy}.  In this paper we explore the effect of ALPS on the propagation of cosmological gamma-rays (see also \cite{DeAngelis:2011id,DeAngelis:2007yu,SanchezConde:2009wu}). In Sec.~\ref{sec:blazar} we will show how ALP-photon oscillations manifest in the observed energy spectra, in Sec.~\ref{sec:calc} we discuss the data sets and proposed analysis technique, and finally we conclude with our results in Sec.~\ref{sec:results}. Averaging over the stochastic intergalactic magnetic field domain structure, we find that the \textit{Fermi} sources can exclude couplings larger than  10$^{-11}$ GeV$^{-1}$ for ALP masses below 10$^{-4} \mu$eV.

\section{ALPs}\label{sec:alps}

ALPs, unlike axions, have a mass and coupling constant that are independent.  A general property of pseudoscalar particles is their connection to the electromagnetic field via a two-photon vertex \cite{1978STIN...7913970D, PhysRevLett.51.1415,Svrcek:2006yi, Masso:2006id}.  Because of this coupling, ALPs and photons can interconvert in the presence of a magnetic field, for example in earth based experiments or, as considered here, in the intergalactic magnetic field (IGMF).  Earth-based exmperiments set constraints on the ALP/axion mass and coupling constant. These include CAST \cite{Zioutas:2004hi, Andriamonje:2007ew} which searches for solar axions, ADMX \cite{Duffy:2006aa} which searches for axion dark matter, and laser regeneration experiments \cite{Chou:2007zzc, Afanasev:2008jt, Zavattini:2005tm}. 

We are concerned with the important feature that to lowest order axions (or more generally, ALPs) can oscillate into photons through the interaction 
\begin{equation}
{\cal L}_{a \gamma} = -\frac{g}{4}~F_{\mu \nu}\tilde{F}^{\mu \nu}a = 
g~{\bf E \cdot B}~a
\label{eq:lagrangiano}
\end{equation}
where $g$ is the photon-axion coupling strength, $a$ is the axion field, $F$ is the electromagnetic field-strength tensor, $\tilde F$ its dual, and \textbf{B} and \textbf{E} are the magnetic and electric field, respectively. This is a three state system: ALP, coupled photon, and decoupled photon. In order for the mixing to occur, there needs to be a coherent magnetic field; the coupled photon parallel to this magnetic field can mix with the ALP state via propogation eigenstates \cite{Csaki:2001yk}. 

Driven by this interaction, photons and ALPs oscillate in and around astrophysical sources with strong magnetic fields as well as in the IGMF and the magnetic field of our own galaxy \cite{Simet:2007sa}.  The IGMF strength is assumed in the range 0.1-1.0 nG \cite{Grasso:2000wj,Dolag:2004kp}.  For a constant magnetic field, the oscillation probabilitiy is given by:
\begin{equation} \label{eq:Pa}
P_o =\frac{1}{1+(E_{crit}/E_{\gamma})^2}~
\sin^2\left[\frac{B~d~g}{2}\sqrt{1+\left(\frac{E_{crit}}{E_{\gamma}}\right)^2}\right]
\end{equation}
where $d$ is the distance traveled and 

\begin{equation}\label{eq:Ecrit}
E_{crit} (GeV) \equiv 2.5 \frac{m^2_{\mu eV}}{B_G~g_{11}}
\end{equation}
Here, the subindices indicate dimensionless quantities: $g_{11}\equiv g/10^{-11}GeV^{-1}$, $m_{\mu eV}\equiv m/\mu eV$, and $B_G\equiv B/Gauss$. The effective ALP mass is defined as: $m^2\equiv \mid m_a^2 - w_{pl}^2 \mid$ and the plasma frequency is: $w_{pl}=0.37\times 10^{-4} \mu eV \sqrt{n_e/cm^{-3}}$ and $n_e\sim 10^{-7}$ cm$^{-3}$. The critical energy is the energy above which oscillation mixing occurs.  For our analysis, we consider parameters so that $E_{crit}$ is at or below the typical energy of the data set (e.g., 30 GeV for \textit{Fermi} and nominally 100 GeV for a very high energy blazar data set).\footnote{In a generic axion model, the mass and coupling are related as $m_a \sim 6 eV ({10^6 GeV}/{f_a})$ where $g=\xi {\alpha}({2 \pi}{f_a})^{-1}$ and $\xi$ is a model dependent factor  of order 1.   Therefore axion-photon oscillations would only occur for $E \geq E_{crit} \sim  10^{18} GeV$ which, for relevant coupling, is far beyond the energies considered in our analysis.}

\begin{figure}
\centering
\includegraphics[height=5cm,width=14cm]{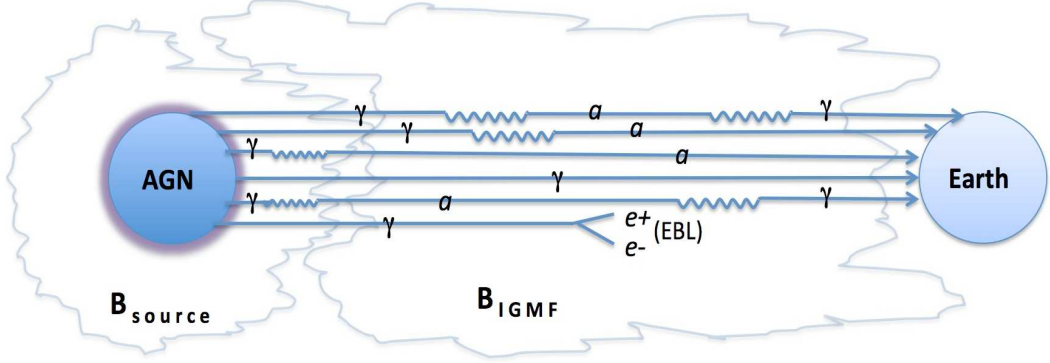}
\caption{\small{Sketch of ALP oscillation scenarios allowing for mixing in/near the source and in intergalactic space. In this paper we only consider mixing in intergalactic space which would eliminate the 3rd and 5th lines from the top. Figure from \cite{SanchezConde:2009wu}.}}
\label{fig:sketch}
\end{figure}

The probability for oscillations is driven by the product $B \times d$.  For oscillations around a source there needs to be a relatively high magnetic field given the relatively short distance scales; this is in contrast to oscillations occurring in the IGMF, which is a much weaker field, but is coherent over  much longer distances. All possible oscillation scenarios are shown schematically in Fig.~\ref{fig:sketch}.  Gamma-ray sources at or above $E_{crit}$ will experience alterations to their energy spectra, an attenuation or enhancement, due to ALP-photon interconversion.  In general, ALP-photon couplings can be probed in a variety of astrophysical and/or cosmological environments provided $B \times d$ is of sufficient magnitude and this quantity is related to the energy of accelerated particles associated with various environments: E$_{max} \simeq 9.3 \times 10^{20}$ eV B$_G$d$_{pc}$, for more see \cite{Hooper:2007bq}. 

Little is understood about the IGMF but it is canonically considered to consist of many coherent domains \cite{0034-4885-57-4-001,0004-637X-556-2-619}, characterized by an average randomly oriented domain length L$_{dom}$ and  $B$  field  strength.
We model the IGMF by taking 1 Mpc and 1 nG (an upperbound IGMF strength \cite{Grasso:2000wj}), respectively.   If the photon beam is taken to be propagating in the $y$ direction then oscillations can occur with $B$ fields in both the $x$ and $z$ directions. The beam can be described by the 3-state vector $({\gamma}_x,{\gamma}_z,a)$ consisting of a coupled photon, a decoupled photon, and an ALP and this beam can interact with the EBL (via pair production with interaction length $l_{EBL} \sim$ 1000 Mpc equivalent to $\tau \sim 0.1$ for HE gamma-rays at z$\sim$1 \cite{Franceschini:2008tp}) while interconverting between photon and ALP. The transfer equation will be \cite{Csaki:2003ef}:

\begin{equation}
\left(\!
\begin{array}{c} 
\gamma_x \\ \gamma_z \\ a
\end{array} 
\!\!\right)
=
{\rm e}^{i E y}
\left[ \, T_0 \, {\rm e}^{\lambda_0 y}
+T_1 \, {\rm e}^{\lambda_1 y} + T_2 \, {\rm e}^{\lambda_2 y} \, 
\right]\!\!
\left(\!
\begin{array}{c}
\gamma_x \\ \gamma_z \\ a                      \label{eq:evolution}
\end{array}
\!\!\right)_{\!\!\!0}   \!
\end{equation}

where:
\begin{eqnarray}
\lambda_0 &\equiv& -\frac{1}{2~{\lambda}_{\gamma}}, \nonumber \\
\lambda_1 &\equiv& -\frac{1}{4{\lambda}_{\gamma}}~\left[ 1 +
\sqrt{1-4~\delta^2} \right] \nonumber \\
\lambda_2 &\equiv& -\frac{1}{4~{\lambda}_{\gamma}}~\left[ 1 -
\sqrt{1-4~\delta^2} \right]
\end{eqnarray}

\begin{eqnarray}
&
T_0 \equiv \left( \begin{array}{ccc}
{\rm sin}^2 \theta & -\, {\rm cos} \theta \, {\rm sin} \theta & 0 \\
-\, {\rm cos} \theta \, {\rm sin} \theta & {\rm cos}^2 \theta & 0 \\
0 & 0 & 0
\end{array} \right) \,\, \qquad
T_1 \equiv \left( \begin{array}{ccc}
\frac{1+\sqrt{1-4\,\delta^2}}{2\,\sqrt{1-4\,\delta^2}}
\, {\rm cos}^2 \theta &
\frac{1+\sqrt{1-4\,\delta^2}}{2\,\sqrt{1-4\,\delta^2}}
\, {\rm cos} \theta \, {\rm sin} \theta &
-\,\frac{\delta}{\sqrt{1-4\,\delta^2}} \,{\rm cos} \theta \\
 \frac{1+\sqrt{1-4\,\delta^2}}{2\,\sqrt{1-4\,\delta^2}}
 \, {\rm cos} \theta \, {\rm sin} \theta &
\frac{1+\sqrt{1-4\,\delta^2}}{2\,\sqrt{1-4\,\delta^2}} \, {\rm sin}^2 \theta &
-\,\frac{\delta}{\sqrt{1-4\,\delta^2}} \,{\rm sin} \theta \\
\frac{\delta}{\sqrt{1-4\,\delta^2}} \,{\rm cos} \theta &
\frac{\delta}{\sqrt{1-4\,\delta^2}} \,{\rm sin} \theta &
-\,\frac{1-\sqrt{1-4\,\delta^2}}{2\,\sqrt{1-4\,\delta^2}}
\end{array} \right) \,\, \nonumber
&
\\
&
T_2 \equiv \left( \begin{array}{ccc}
-\,\frac{1-\sqrt{1-4\,\delta^2}}{2\,\sqrt{1-4\,\delta^2}} \, {\rm cos}^2 \theta &
-\,\frac{1-\sqrt{1-4\,\delta^2}}{2\,\sqrt{1-4\,\delta^2}}
\, {\rm cos} \theta \, {\rm sin} \theta &
\frac{\delta}{\sqrt{1-4\,\delta^2}} \,{\rm cos} \theta \\
-\,\frac{1-\sqrt{1-4\,\delta^2}}{2\,\sqrt{1-4\,\delta^2}}
\, {\rm cos} \theta \, {\rm sin} \theta &
-\,\frac{1-\sqrt{1-4\,\delta^2}}{2\,\sqrt{1-4\,\delta^2}}
\, {\rm sin}^2 \theta &
\frac{\delta}{\sqrt{1-4\,\delta^2}} \,{\rm sin} \theta \\
-\,\frac{\delta}{\sqrt{1-4\,\delta^2}} \,{\rm cos} \theta &
-\,\frac{\delta}{\sqrt{1-4\,\delta^2}} \,{\rm sin} \theta &
\frac{1+\sqrt{1-4\,\delta^2}}{2\,\sqrt{1-4\,\delta^2}}
\end{array} \right) ~
&
\end{eqnarray}
where $\theta$ is the angle between the $x$-axis and the direction of B in each single domain and $\delta$ is a dimensionless parameter equal to:
\begin{equation}
\label{delta}
\delta \equiv B {g}{l}_{EBL}  \simeq 0.11 \left( \frac{B}{10^{-9}\, {\rm G}} \right) \left( \frac{g}{10^{-11} \, {\rm GeV}} \right)
\left( \frac{{l}_{EBL}}{{\rm Mpc}} \right)
\end{equation}
which for our calculation is of order 100.  In this limit, there is a simple and intuitive analytic calculation for a constant B field that we also consider. For $\delta \>>$ 1 \cite{Csaki:2003ef},

\begin{eqnarray}
I_\gamma  &\simeq& \frac{1}{2} \, {\rm e}^{-\frac{y}{l_{\rm EBL}}} +
\frac{1}{2} \, {\rm e}^{-\frac{y}{2\,l_{\rm EBL}}} \, {\rm cos}^2 \left(
\frac{\delta y}{2\,l_{\rm EBL}} \right) \nonumber\\
\label{eq:constB}
\end{eqnarray}  
The first term corresponds to the decoupled photon which can only interact with the EBL. The second term corresponds to the coupled photon-ALP system which we can view as photon-like half of the time and ALP-like half of the time. Thus the mean free path of interactions with the EBL is doubled. 

\begin{figure}
   \centering
   \includegraphics[width=0.65\textwidth]{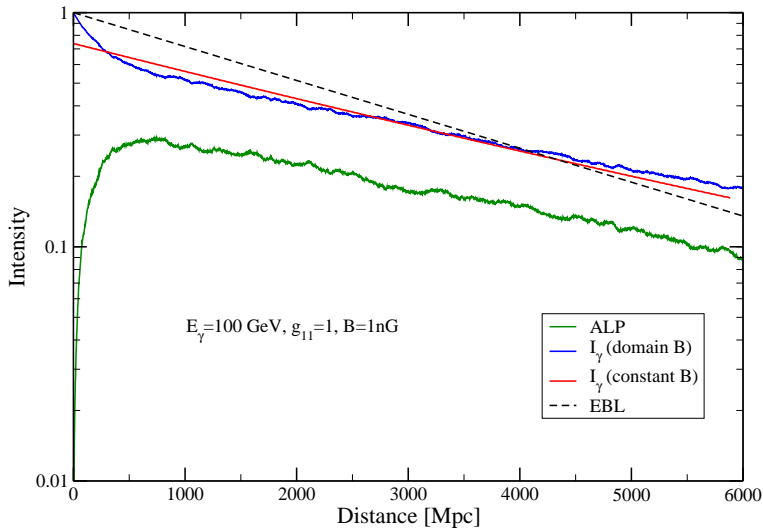}

\caption{The photon intensity for both the varying B field formalism and the constant B field formalism. The ALP intensity and EBL-only line are shown for comparison. }
\label{fig:compare}
\end{figure}

In general the random orientation of the magnetic domains leads to a stochastic photon intensity as a function of distance from a single source, with a length scale comparble to the domain size and average fluctuation amplitude  $\sim 0.3$.  We can average over many realizations of the domain alignment (i.e., lines of sight to many blazars in a data set) which reduces the stochasticity and corresponds to the average ALP effect as obtained for a given data set.    In Fig.~\ref{fig:compare} we show how the predictions of both the constant B field and average over a 1000 realizations of the small domain magnetic field compare for a set of parameters relevant to this paper. For the distance scale we are considering ($z\sim 1$ or $d\sim$ 5000 Mpc), this difference is neglibible given the analysis technique we imploy. For the upcoming calculation we will use the intuitive and simple constant B field formalism.  In addtion, we show the ALP intensity for the domain averaged calculation.  The physical interpretation is fairly simple: on average, photons and ALPs readily interconvert on scales short compared to the source distance while photons are depleted by pair production off the EBL on intermediate scales, leading to a decreasing photon and ALP intensity (with roughly constant ratio) with distance. For comparison, we show the EBL pair production normalized to produce the Fermi data set.  The particular ALP solution shown corresponds to an enhancement to the photon flux at 5000 Mpc (relative to the pure EBL) due to ALPs converting to photons within the EBL horizon.  There is an analogous ALP solution corresponding to extra attenuation of the photons (again, relative to the pure EBL) due to a slightly small mixing of ALPs, allowing for more photon attenuation near the source prior to the full population of the ALP state.

\section{Blazar Spectra}

\label{sec:blazar}
We know that gamma-rays coming from distant blazars interact with the EBL, creating electron-positron pairs, and attenuating the energy spectrum  \cite{Ackermann:2012sza, Reesman:2013jra}. This attenuation corresponds to an optical depth $\tau=\tau(E,z)$ such that $F_{obs}=F_{s}e^{-\tau}$, where $F_s$ is the energy spectrum at the source.
If the EBL and source spectra were known precisely, then the residual between the predicted and observed spectra could be used to constrain other photon interactions beyond the standard model.  In practice, neither the EBL nor the source spectra are known to a high enough level of accuracy to meaningfully constrain new physics.\footnote{Pseudoscalars like axions and ALPs, can also be produced in the accretion disk of active galactic nuclei by Compton, Bremsstrahlung, and Primakoff processes, though it has been calculated to be negligible compared to the photon luminosity for g$_{a \gamma\gamma} =$ 8.4$\times$ 10$^{-12}$ GeV$^{-1}$ \cite{Jain:2009hf}.}    Never-the-less, we can pick a well-fit EBL model using all existing data (see Fig. 1 in \cite{Ackermann:2012sza} and Fig. 2 in \cite{Reesman:2013jra}) and assign the errors in that fit to potential new physics in order to illustrate the possible constraints achievable.   Here we will assume the theoretical EBL model given by \cite{Franceschini:2008tp} to be correct, and allow the error in the best-fit  $\tau$ to be due to ALP physics. This allows us to place conservative constraints on ALP-photon coupling and mass. 

We rewrite the spectrum as:

\begin{equation}
F_{obs}=F_{s}e^{-\tau}\phi_{ALPs},
\end{equation} 
where $\phi_{ALPs}=e^{\pm\bigtriangleup\tau}$. The plus sign indicates there was less attenuation than predicted by the model and thus gamma-rays are \textit{enhanced}. Conversely, the negative sign indicates there was more attenuation than predicted by the model and thus extra \textit{extra attenuation}. 

We can use Eq.~\ref{eq:constB} to describe the observed flux, F$_{obs}$=F$_{s}I_{\gamma}$. Then we can rewrite it by pulling out the term specific to the EBL: 

\begin{equation}
F_{obs}=F_{s}e^{-\tau}\left(\dfrac{1}{2} + \dfrac{1}{2}e^{\tau /2}~cos^2(\dfrac{\tau}{2}~\delta)\right)
\end{equation}
since $y$ now corresponds to the total distance traveled we have replaced $y/l_{EBL}$ with $\tau$ . 

This means that 
\begin{equation}
\phi_{ALPs}=e^{\pm\bigtriangleup\tau}=\dfrac{1}{2} + \dfrac{1}{2}e^{\tau /2}~cos^2(\dfrac{\tau}{2}~\delta)
\label{eq:ALPpiece}
\end{equation}

Thus for a given $\bigtriangleup\tau$ we can find a best fit $\delta$ and best fit coupling.

\section{Calculations} \label{sec:calc}
At present, attenuation due to the EBL is not precisely known at any energy. What is known is that a suite of reasonable models provide a good fit to the EBL opacity for photons in the energy range 50 GeV - 10 TeV \cite{Reesman:2013jra}. The error in the EBL model fit is driven primarily by the uncertainty in the intrinsic spectrum produced by blazars. This can be seen by comparing the 1 and 2 $\sigma$ regions of the optical depth given in \cite{Ackermann:2012sza} to \cite{Reesman:2013jra}.  Never-the-less we can illustrate the expected range of applicable ALP physics by setting the absorption/amplification due to ALP-photon interconversion to be no larger than the error in the best fit optical depth.  That is to say, EBL models provide a fairly accurate description of the absorption observed in the current sample of gamma-ray blazars.  What ALP physics there is should not be so large as to spoil this agreement.

\begin{figure}
\centering
\includegraphics[width=0.60\textwidth]{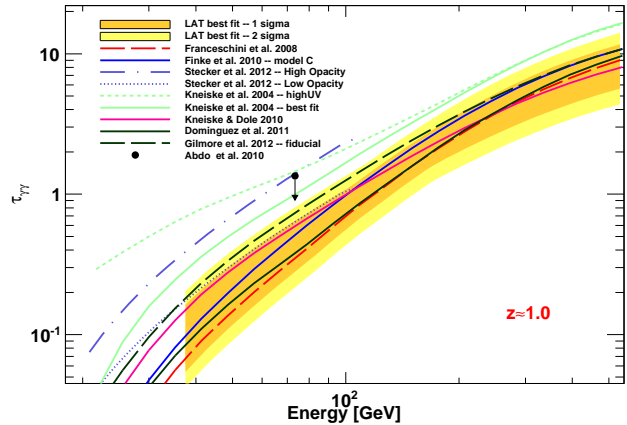}
\caption{The opacity of the universe to high energy gamma-rays, as determined by \textit{Fermi}.  Figure from \cite{Ackermann:2012sza}. }\label{fig:fermi_tau}
\end{figure}

Our analysis was done on the results presented by \textit{Fermi} \cite{Ackermann:2012sza} on the optical depth of the Universe as measured by a set of 150 blazars, see Fig.~\ref{fig:fermi_tau}. For their analysis the blazars were split into three redshift bins with values ranging from $ 0.03 < z < 1.6$. The middle bin had the strongest constraining power and constituted an average redshift, $z\sim 1.0$ (or an average source distance of $\sim 5000$ Mpc). They treated the values of the energy spectrum below $E<25$ GeV as the true, unabsorbed, intrinsic source spectrum. Above this value the spectrum was assumed to interact with the EBL. For our analysis we consider E$_{crit} \le 30 GeV$ so that ALP-photon interactions occur for all energies probed by \textit{Fermi}. This choice for E$_{crit}$ thus avoids threshold issues \cite{Wouters:2012qd}  and makes oscillations energy independent for E $>$ E$_{crit}$.  Assuming perfect knowledge of the EBL, as described in Sec.~\ref{sec:blazar}, we can put conservative constraints on ALP parameters from this data set. 

\begin{figure}
 \centering
  \includegraphics[width=0.60\textwidth]{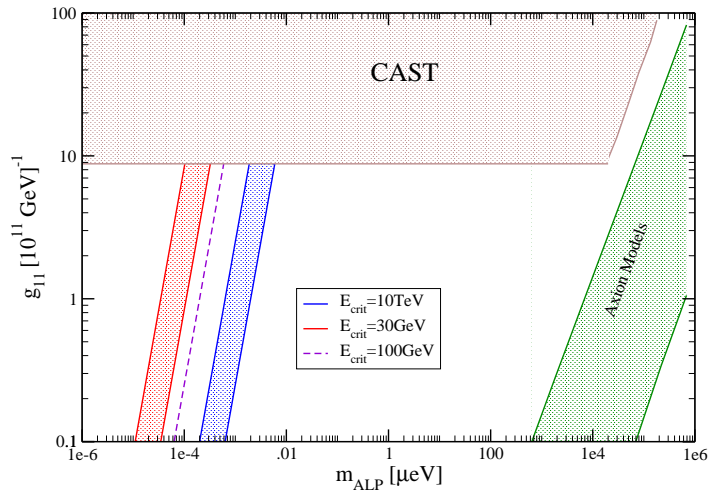}
 \caption{The expected region of significant ALP conversion as a function of the coupling and ALP mass for data at the indicated redshifts.  The red and blue bands correspond to critical energies of the indicated value. The spread in each band accounts for different $B$ field values ranging from $0.1-1$ nG.}
 \label{fig:paramspace}

\end{figure}

In order to constrain ALP physics from a given source, the critical energy needs to be at or below the range of gamma-ray energies produced by the source. Otherwise, no oscillations occur for some or all of the photons emitted by the source. Limiting the amount of attenuation (either for \textit{enhancement} or \textit{extra attenuation}) leads to a constraint on $\delta$ and the coupling strength which in turn constrains the mass, for a given E$_{crit}$. 
In Fig.~\ref{fig:paramspace}, the red (blue) slanted region corresponds to the allowed paramter space for $E_{crit}=30$  GeV ($E_{crit}=10$ TeV). The width corresponds to the allowed range in $B$ field.

\section{Results and Discussion} \label{sec:results}
We have shown how to probe the scale of ALPs coupling to high energy photons if we assume that the residual between the observed optical depth and a best fit model (\textit{Fermi} collaboration \cite{Ackermann:2012sza}), cannot be exceeded by the ALP-photon interconversion associated with this coupling. For a detailed calculation of strong mixing in the IGMF see the DARMA model in \cite{DeAngelis:2011id,DeAngelis:2007yu}.

\begin{figure}
 \centering
 \includegraphics[width=0.65\textwidth]{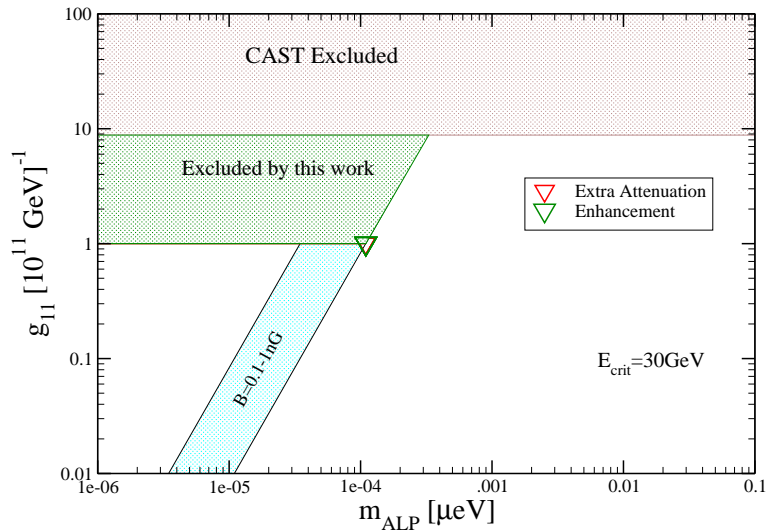}
 \caption{ALP photon coupling and ALP mass  as constrained by HE blazar observations made by the Fermi collaboration. CAST line from \cite{Andriamonje:2007ew}.}
 \label{fig:constraint}
\end{figure}

In Fig.~\ref{fig:constraint} we illustrate the results of this method.  As in Fig.~\ref{fig:paramspace}, the slanted region corresponds to the paramter space such that $E_{crit}=30$ GeV and  the spread is for a range of IGMF $B$ field values. The actual calculation was done for $B=1$ nG, causing the maximal allowed ALPS coupling to lie on the right side of the 30 GeV band. Had we used a smaller magnetic field value (for example, $B=0.1$ nG) then the upper-bound of the mass would be a factor of three smaller for the same E$_{crit}$. The two constraint points shown correspond to the exact limit from the \textit{Fermi} data assuming a critical energy of 30 GeV.  Moving to lower $m_{\mu eV} $ corresponds to lower values of E$_{crit}$, which would still allow all the photons from the source to oscillate. Therefore points to the left of the shown constraints result in the same ALP distortion.  Similarly, points below the data constraints and along the lines of constant E$_{crit}$ correspond to longer decay lengths (lower probabilities for conversion) and are therefore allowed. The green shaded region is excluded by this data, under our assumptions. The red shaded region corresponds to points that are acceptable for \textit{enhancement}, but unacceptable for \textit{extra attenuation}. Both solutions correspond to roughly the same coupling, with the \textit{enhancement} solution for slightly larger coupling. \footnote{Back conversion in the Milky Way was used to constrain ALP-photon coupling using the lack of 100 MeV gamma rays from SN87a \cite{Simet:2007sa}.  In this analysis, high energy ALPS are produced in the hot core of the supernova and then back convert in the Galaxy's magnetic field.  The absence of high energy gamma rays from SN87a then constrains $g_{11} \geq 1$ for $m_{\mu eV} <  10^{-3}$.  For the parameter range considered here, back conversion of blazar ALPs to photons would also occur in our Galaxy, but at a level minimal compared to the amplitude of residuals we consider.}

Historically, the effect of ALPs on blazars has focused on examining the spectrum of an individual source and attributing a break in the assumed source power-law to ALPs. These analyses run Monte Carlo simulations over ALP parameter space, including a parameterization of magnetic field domains, among other quantities \cite{ Tavecchio:2012um, SanchezConde:2009wu}. Our method instead illustrates the range of constraints possible if we integrate over the spectral variations of many sources, at approximately the same distance, at the same time. This is completely analogous to the EBL measurement made by \textit{Fermi} and hence why we can use their result.  In a perfect world, improving the data set for TeV blazars at $z\sim 0.1$ could be used in concert with the \textit{Fermi} blazars to better probe slightly larger ALP masses at roughly the same coupling, in particular in the region where both experiments are measuring the same EBL \cite{Reesman:2013jra}.  

In \cite{SanchezConde:2009wu}, ALP mixing in both the source and the IGMF was considered for 3C 279 (z$=$0.536) and PKS 2155-304 (z$=$0.117). Their fiducial model used the values g$_{11}=8.77$ and m$_{\mu eV}=10^{-4}$. ALP-photon mixing can also be used to reduce the photon opacity within the source itself -- in a recent paper \cite{Tavecchio:2012um} the possibility of ALP-photon conversion in the source PKS 1222+216  (z$=$0.432) was explored as a possible explanation of reducing the source opacity. They considered the presence of a strong B field in/around the source that allowed for $\gamma\rightarrow a$ conversion such that the gamma-ray could escape the broad line region of the blazar. Once escaped, the ALP could convert back $a\rightarrow \gamma$ in either the host galaxy, or in the IGMF. In order to explain both the HE and VHE observations (from \textit{Fermi} and MAGIC) using a standard blazar model, the authors needed to invoke ALPs with an inverse coupling g$_{11}=$1.4 in a (source) magnetic field $B_{G}=0.2$.

That is to say, if ALPs are required to extend the TeV horizon  and/or reduce the source opacity of several 100 GeV photons in blazars, we would expect correlated ALP effects, at a scale demonstrated in this paper, in the set of \textit{Fermi} blazars clustered near redshift 1.

\bibliographystyle{plain}
\bibliography{ALP_bib_long}

\end{document}